\documentclass[12pt,preprint]{aastex}

\begin{document}
  \title{Irradiated ISM: Discriminating between Cosmic Rays and X-rays}

  \author{R. Meijerink} \affil{Sterrewacht Leiden, P.O. Box 9513, 2300
  RA, Leiden, The Netherlands}\email{meijerin@strw.leidenuniv.nl}

  \author{M. Spaans} \affil{Kapteyn Astronomical Institute, P.O. Box
  800, 9700 AV Groningen, The Netherlands}\email{spaans@astro.rug.nl}

  \and

  \author{F.P. Israel}\affil{Sterrewacht Leiden, P.O. Box 9513, 2300
  RA, Leiden, The Netherlands}\email{israel@strw.leidenuniv.nl}

  \begin{abstract}
  The ISM of active galaxy centers is exposed to a combination of
  cosmic ray, FUV and X-ray radiation. We apply PDR models to this ISM
  with both `normal' and highly elevated ($5\times 10^{-15}$~s$^{-1}$)
  cosmic-ray rates and compare the results to those obtained for
  XDRs. Our existing PDR-XDR code is used to construct models over a
  $10^3-10^5$ cm$^{-3}$ density range and for 0.16-160 erg s$^{-1}$
  cm$^{-2}$ impingent fluxes. We obtain larger high $J$ ($J>10$) CO
  ratios in PDRs when we use the highly elevated cosmic ray rate, but
  these are always exceeded by the corresponding XDR ratios. The [CI]
  609 $\mu$m/$^{13}$CO(2-1) line ratio is boosted by a factor of a few
  in PDRs with $n\sim 10^3$ cm$^{-3}$ exposed to a high cosmic ray
  rate. At higher densities ratios become identical irrespective of
  cosmic ray flux, while XDRs always show elevated [CI] emission per
  CO column. The HCN/CO and HCN/HCO$^+$ line ratios, combined with
  high $J$ CO emission lines, are good diagnostics to distinguish
  between PDRs under either low or high cosmic ray irradiation
  conditions, and XDRs. Hence, the HIFI instrument on Herschel, which
  can detect these CO lines, will be crucial in the study of active
  galaxies.
  \end{abstract}

  \keywords{ISM: X-rays --- ISM: Cosmic Rays}
%

\section{Introduction}

In centers of late-type galaxies, such as M~82, NGC~253, and Maffei 2,
molecular line intensity ratios\footnote{Intensity ratios of lines a
to lines b are related to brightness temperature ratios by the cube of
the line frequencies: ($\nu_{a}/\nu_{b})^{3}$} are frequently found to
be high, e.g. CO(4-3)/CO(1-0)$\sim 60$ \citep{Israel1995, White1994,
Israel2003} requiring high gas densities $n>10^{5.5}$~cm$^{-3}$, and
high temperatures of $T\geq50$~K. However, FUV photons are easily
attenuated by dust and do not penetrate very deep into clouds and the
galaxies are observed with beams covering regions typically hundreds of
parsecs in size.  At such large spatial scales, excited dense gas is
not expected to have very large filling factors. UV radiation seems
incapable of maintaining very large gas fractions at temperatures of
$T\sim50-150$~K, and regular PDR models do not explain the observed
high ratios.  In addition, \citet{Israel2002} have observed the [CI]
609~$\mu$m line in the centers of late-type galaxies, and measured
[CI]~609~$\mu$m/$^{13}$CO(2-1) line intensity ratios in the range 20
-- 60, which are hard to explain by (low-CR) PDR models. The
C$^+$/C/CO transition zones in these PDRs contain only a thin layer in
which neutral carbon has a high abundance. At the cloud edge, carbon
is ionized, and deep in the cloud all carbon is locked up in CO.

Various authors have invoked elevated cosmic ray fluxes caused by
greatly enhanced supernova rates in galaxy centers in order to explain
large molecular gas masses at high temperatures
\citep{Suchkov1993,Bradford2003}, and to explain high [CI] intensities
and column densities relative to $^{12}$CO and $^{13}$CO
\citep{Pineau1992, Schilke1993, Flower1994}. As most cosmic rays are
produced in supernovae, their flux is proportional to the star
formation rate, which is about $1$~M$_\odot$~yr$^{-1}$ for the Milky
Way.  In circumnuclear starbursts, star formation rates may be two
orders of magnitude higher or more.  Such galaxy centers may also
contain an embedded accreting black hole producing X-rays.  Like
cosmic rays, but unlike UV photons, these X-rays can also penetrate
through large column densities ($N_H>10^{24}$~cm$^{-2}$), and can
cause the observed high line ratios over areas as large as 500~pc,
when the emitted flux is high enough \citep{Meijerink2005, Meijerink2006}.

In this paper, we investigate whether in galaxy central regions PDRs
with and without enhanced cosmic ray fluxes can be distinguished from
XDRs on the basis of observable atomic and molecular line ratios. To
this end, we calculate line intensities for PDRs with very different
cosmic ray rates and compare the results to those obtained for XDRs 
with the same radiation fields and column densities.

\section{PDR and XDR models}

We have constructed a set of PDR and XDR models from the codes
described by \citet{Meijerink2005} and \citet{Meijerink2006}, in which
we varied both the incident radiation field and the density. The
thermal balance (with line transfer) is calculated self-consistently
with the chemical balance through iteration. Absorption cross sections
for X-rays (1-100 keV) are smaller, $\sim 1/E^3$, than for FUV
photons. Therefore, PDRs show a stratified structure while the changes
in the chemical and thermal structure in XDRs are very
gradual. Species like C$^+$, C and CO co-exist in XDRs, and large
columns of neutral carbon (unlike in PDRs) are produced. In the XDRs,
additional reactions for fast electrons that ionize, excite and heat
the gas are included. The heating efficiency in XDRs is much
higher. Since we focus on galaxy centers, we have assumed the
metallicity to be twice Solar. We take the abundance of carbon to be
equal to that of oxygen, since the carbon abundance increases faster
than oxygen for larger metallicity. The precise C:O ratio does not
affect our general results. We have calculated PDR models for both a
`normal' (low: $\zeta=5\times10^{-17}$~s$^{-1}$ -- cf. van der Tak
$\&$ van Dishoeck \citeyear{vdTak2000}) and a high
($\zeta=5\times10^{-15}$~s$^{-1}$) cosmic ray flux corresponding to a
star formation rate of $\sim 100$ M$_\odot$ yr$^{-1}$. The model input
parameters are summarized in Table \ref{models}. The range of free
parameters ($n=10^3-10^5$~cm$^{-3}$,
$G_0=10^2-10^5$/$F_X=0.16-160$~erg~s$^{-1}$~cm$^{-2}$) is
representative for the conditions in galaxy centers. That is, from
low-$J$ CO (critical density $10^3-10^4$ cm$^{-3}$) and HCN 1-0
(critical density $\sim10^5$ cm$^{-3}$) observations it is apparent that
gas in galaxy centers must exhibit the density range that we model.
The irradiation conditions are typical for Milky Way PDRs like the
Orion Bar (a high mass star-forming region) as well as a generic
$10^{44}$ erg s$^-1$ Seyfert nucleus X-ray luminosity for distances of
about 100 pc and up.

\section{Chemical and thermal structure}

\begin{figure}
\begin{center}
\unitlength1cm
\begin{minipage}[b]{5.25cm}
\resizebox{5.5cm}{!}{\includegraphics*[angle=0]{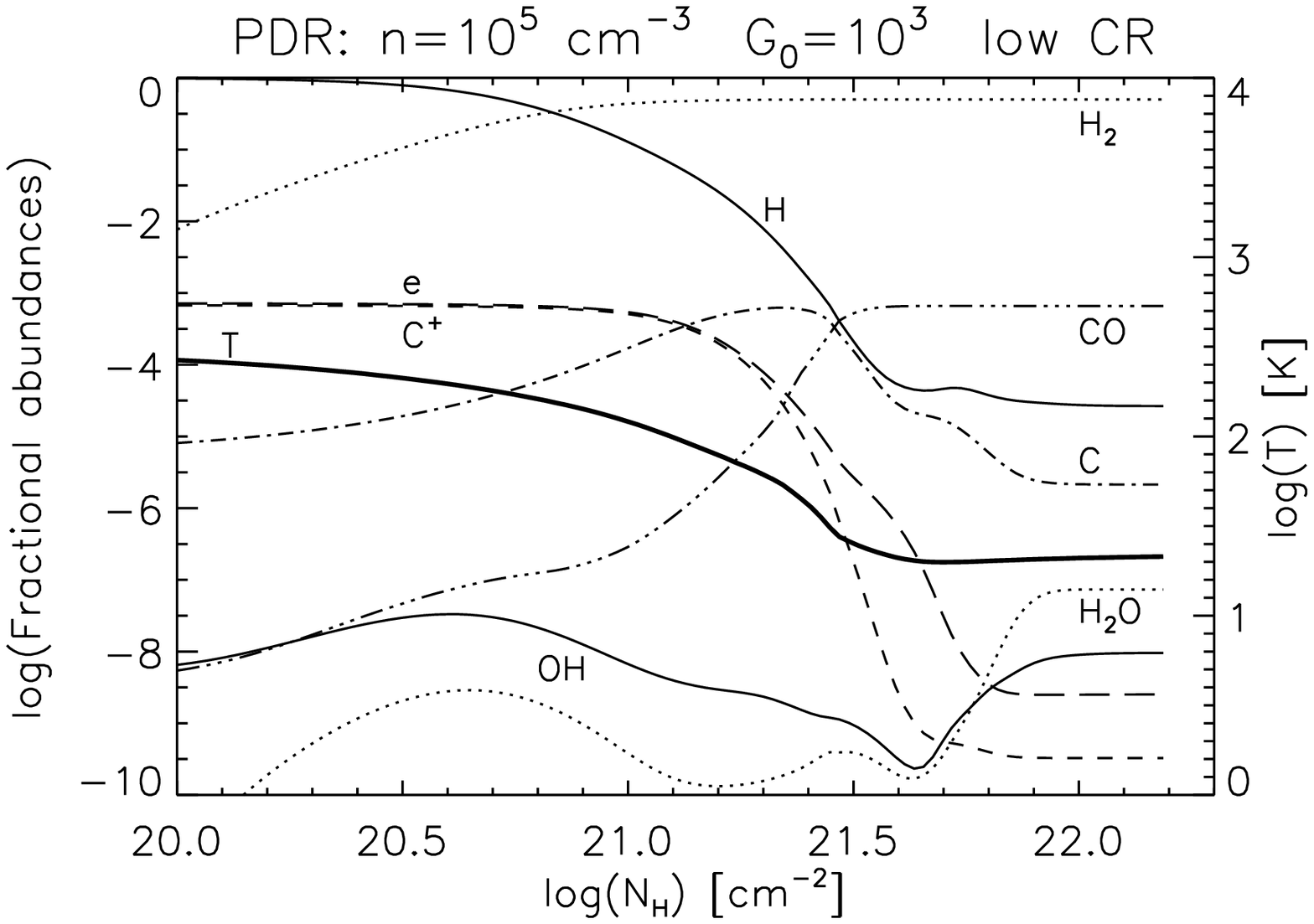}}
\end{minipage}
\begin{minipage}[b]{5.25cm}
\resizebox{5.5cm}{!}{\includegraphics*[angle=0]{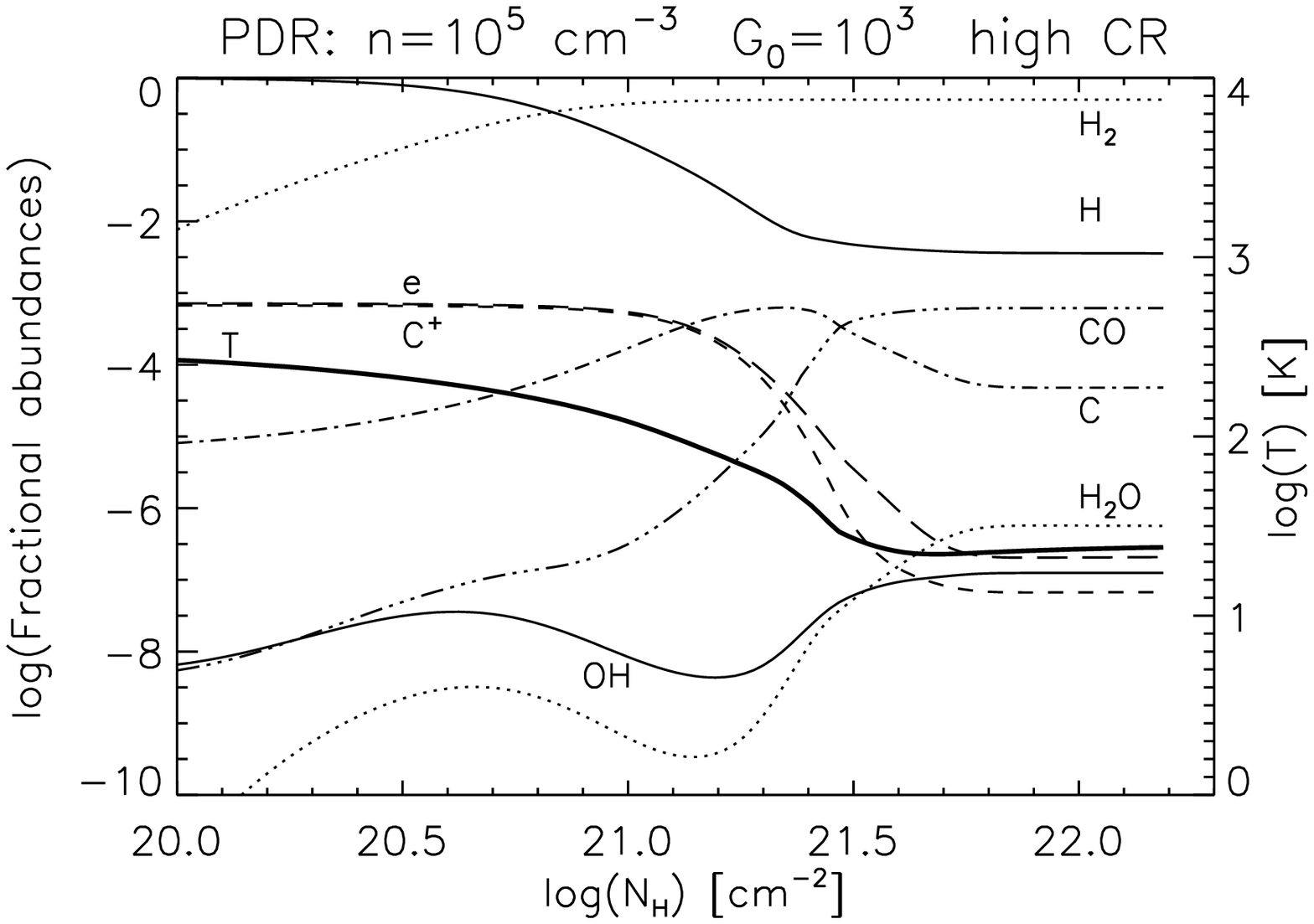}}
\end{minipage}
\begin{minipage}[b]{5.25cm}
\resizebox{5.5cm}{!}{\includegraphics*[angle=0]{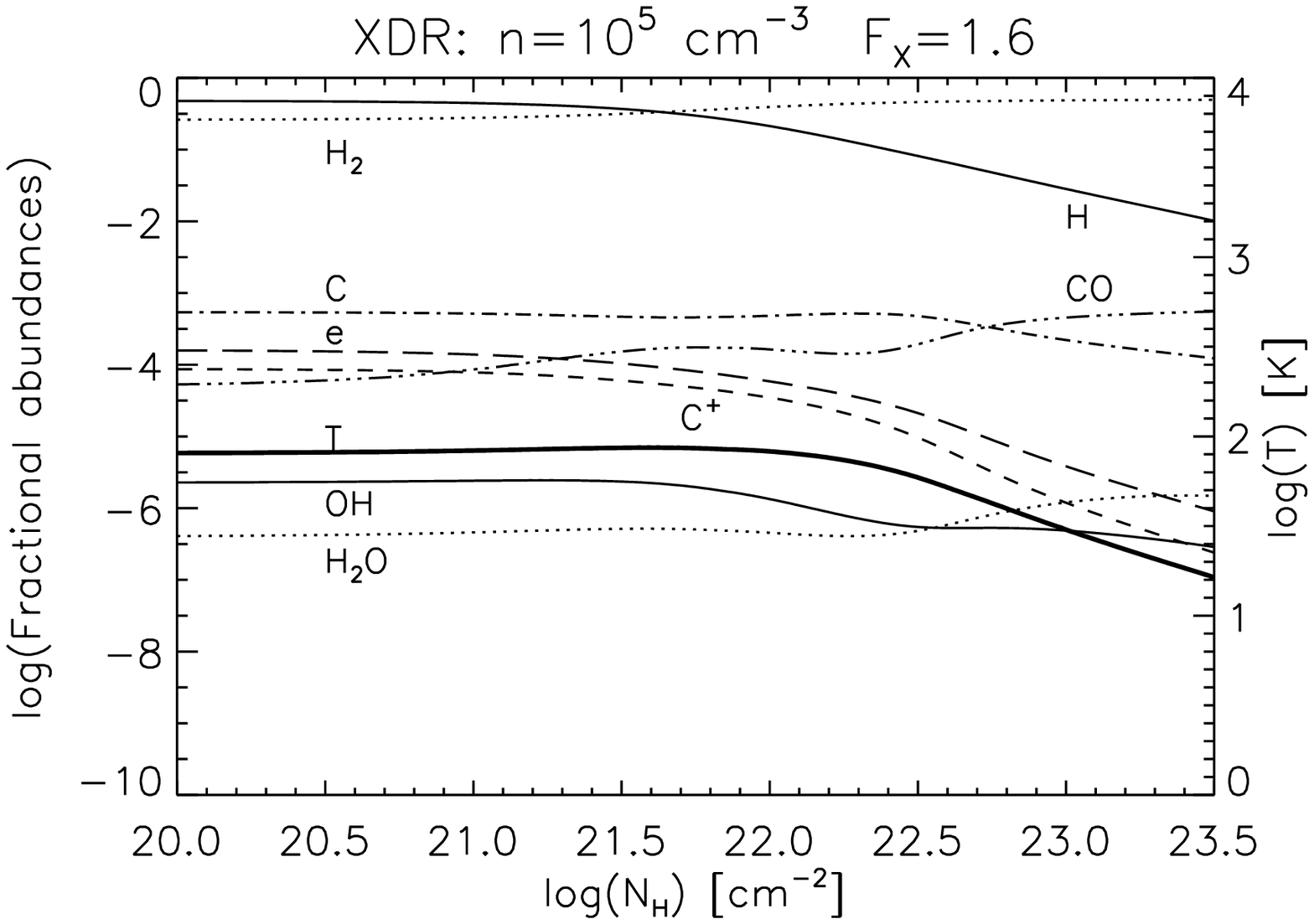}}
\end{minipage}
\begin{minipage}[b]{5.25cm}
\resizebox{5.5cm}{!}{\includegraphics*[angle=0]{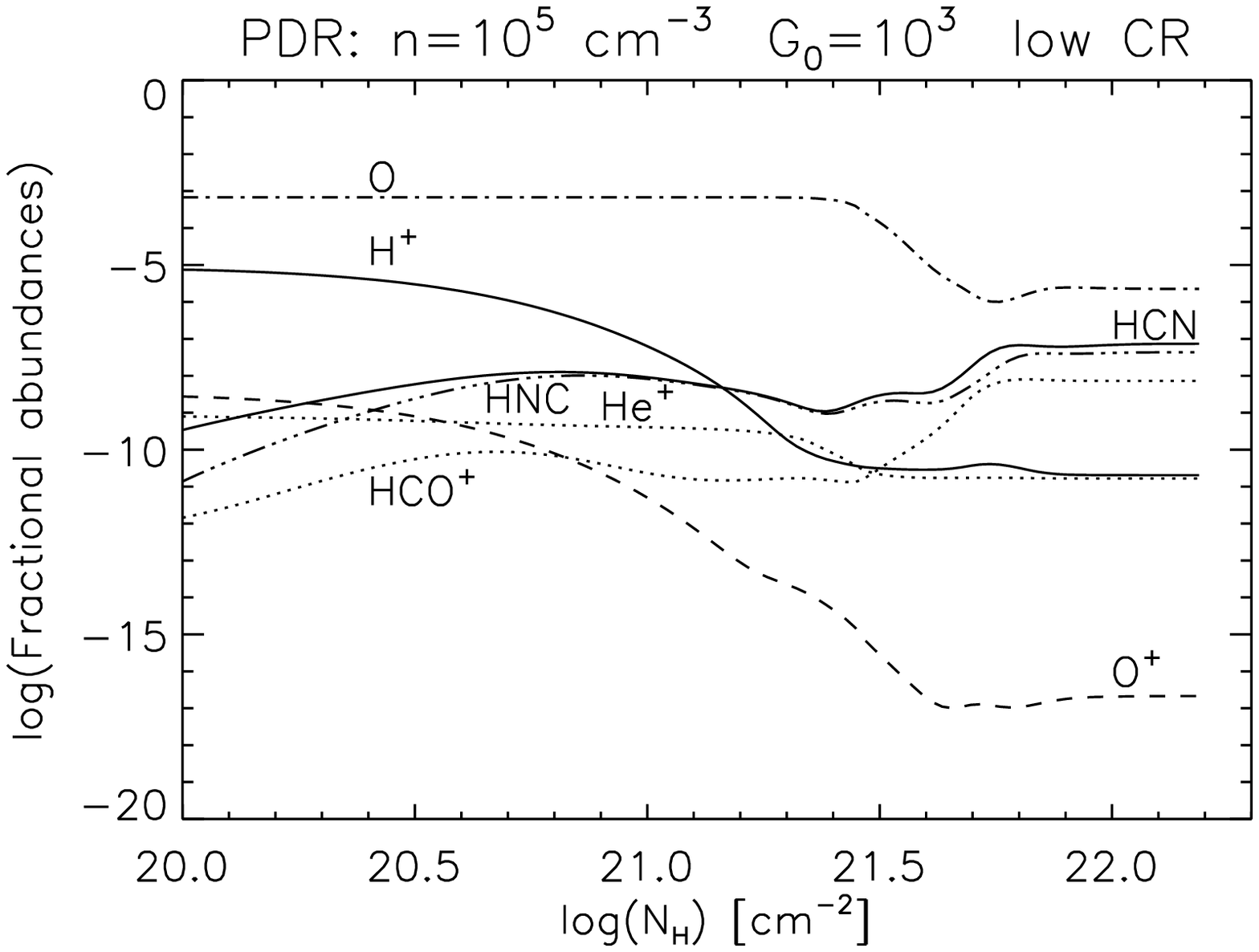}}
\end{minipage}
\begin{minipage}[b]{5.25cm}
\resizebox{5.5cm}{!}{\includegraphics*[angle=0]{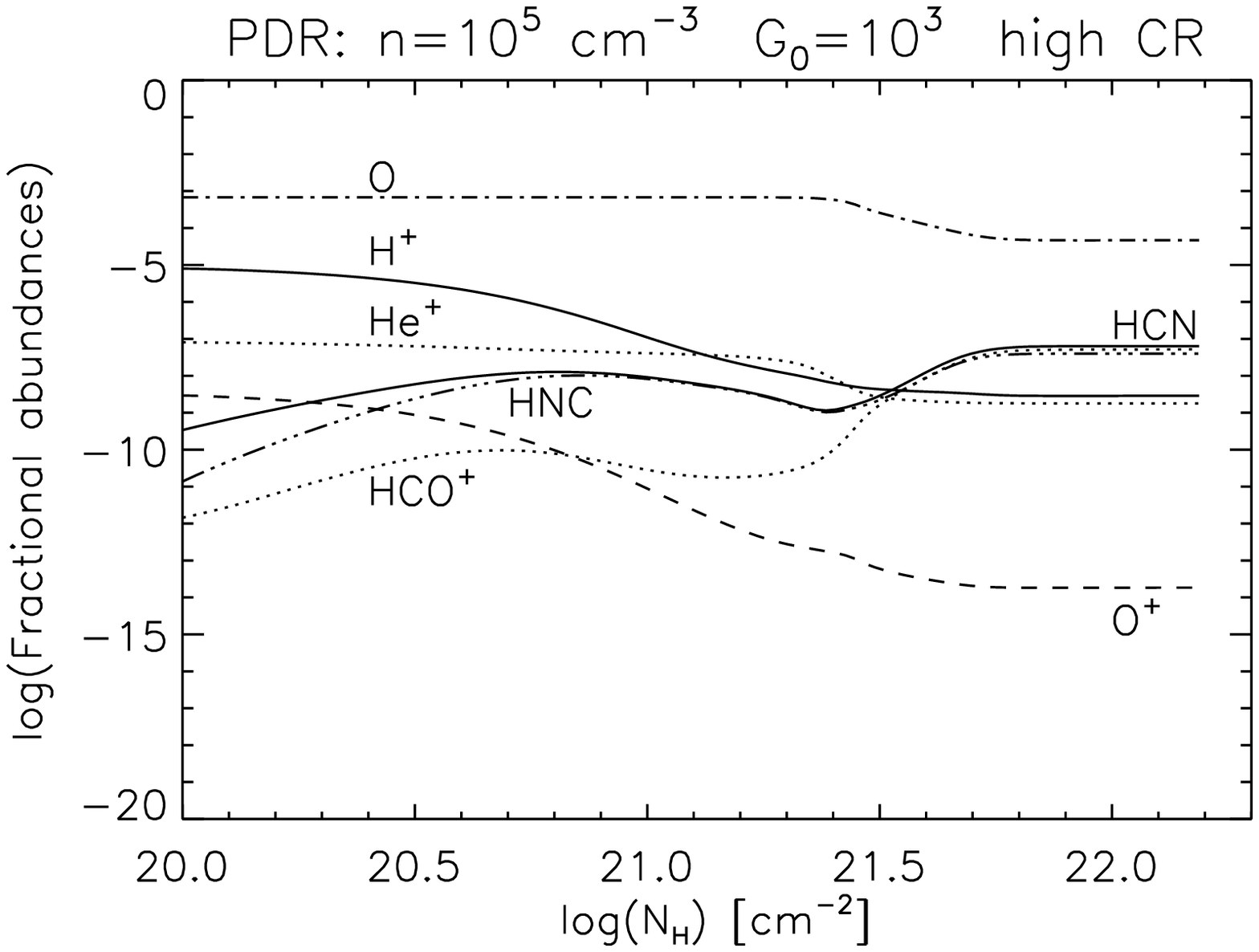}}
\end{minipage}
\begin{minipage}[b]{5.25cm}
\resizebox{5.5cm}{!}{\includegraphics*[angle=0]{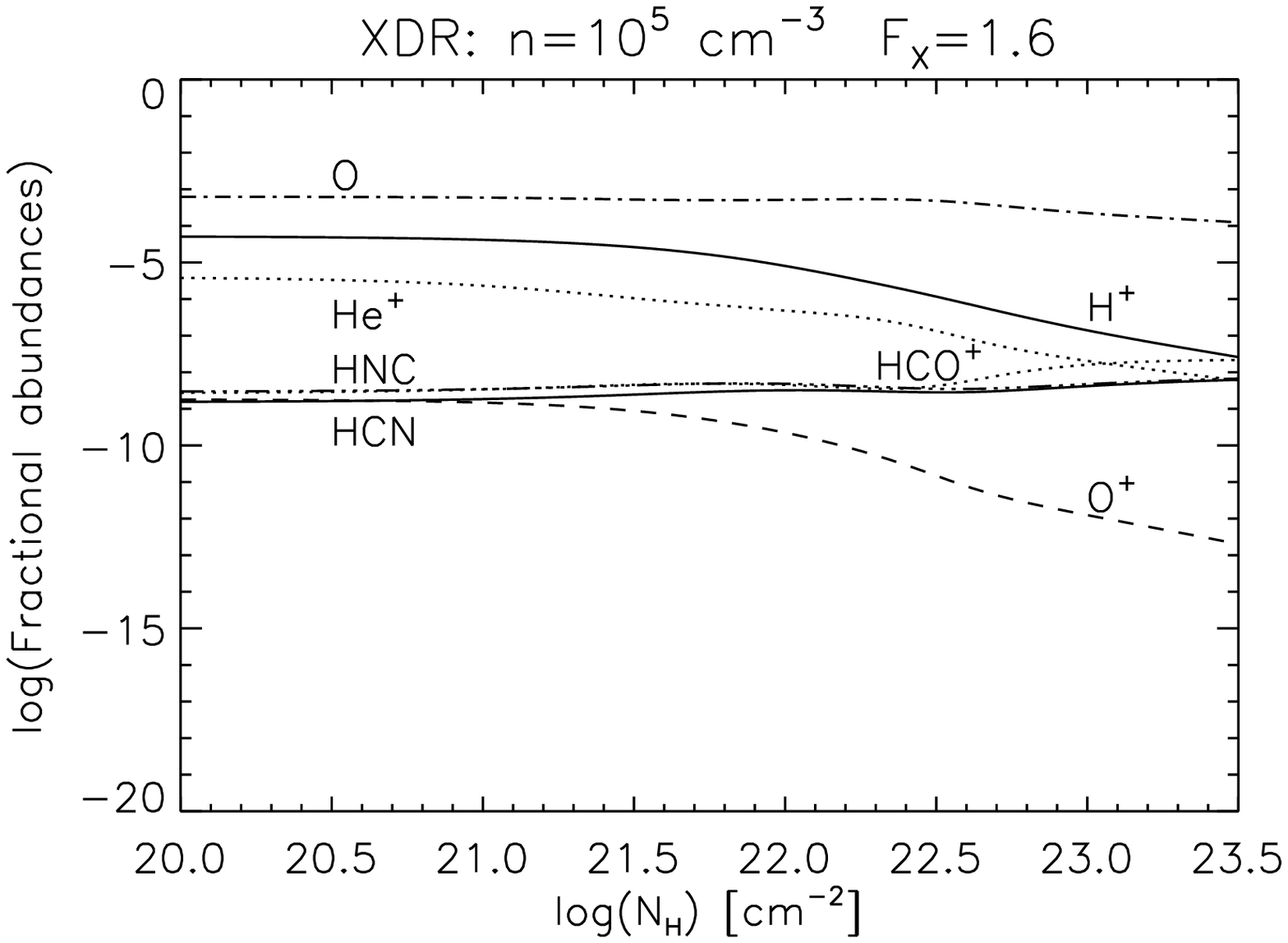}}
\end{minipage}
\caption[] {Chemical and thermal structure of PDR and XDR models at
density $n=10^5$~cm$^{-3}$ and $G_0=10^3$
($F_X=1.6$~ergs~s$^{-1}$~cm$^{-2}$)}
\label{PDRandXDR}
\end{center}
\end{figure}

Higher cosmic ray (CR) ionization rates do not much affect the
chemistry at the cloud edge, but large effects occur beyond the
H/H$_2$ transition (see e.g. Fig. \ref{PDRandXDR}).  As the CR flux in
the PDRs is enhanced, electron, carbon and hydrogen abundances
decrease much less beyond the H/H$_2$ transition, due to larger
ionization/dissociation rates.  Ion abundances also remain higher,
causing H$_2$O and OH to have higher abundances as well.  As cosmic
ray ionization contributes to the gas heating, higher incident CR
fluxes raise gas temperatures deep in the cloud, with roughly
$T\sim\zeta^{1/3}$ at low densities ($n\sim 10^3$~cm$^{-3}$). At high
densities, $n=10^5$~cm$^{-3}$, dust acts as an effective coolant if
$T > T_d$, and the rise in the kinetic temperature is less
pronounced. Both temperatures and abundances in a PDR become higher
when CR rates are increased and ratios of emergent line emissions are
also modified.  

\section{CO line intensities and ratios}

{\it a. Density $n=10^3$~cm$^{-3}$} (Table \ref{low_dens}). In high-CR
PDR clouds, a large fraction of all CO is dissociated, and CO is a
factor $\sim100$ less abundant than in low-CR PDRs. However, in the
latter, the CO lines are optically thick and CO line intensities are
comparible for the same incident flux.  In high-CR PDRs, the low-$J$
CO line ratios are somewhat larger than those in low-CR PDRs. The
higher transitions show more of a difference and are diagnostically
more valuable.  In the XDRs, the CO gas temperature is on average much
higher than in the PDRs, but marginally higher CO intensities occur
only for the weak radiation field $F_X=1.6$~ergs~s$^{-1}$~cm$^{-2}$.
{\it In XDRs with stronger radiation fields, CO line intensities are
much lower than in PDRs}, because column densities are too small to
attenuate the X-rays significantly, and CO abundances are very low
throughout the cloud.  The very high CO(2-1)/CO(1-0) and
CO(4-3)/CO(1-0) ratios in XDRs, which can be much higher than in the
corresponding PDR cases, merely reflect the weakness of the lower $J$
lines.

{\it b. Density $n=10^4$~cm$^{-3}$} (Table \ref{mid_dens}). High-CR
PDRs have more intense CO lines than low-CR PDRs, but {\it their
high-$J$/low-$J$ CO line ratios} are only marginally higher than those
of the corresponding low-CR PDRs.  Again we find that XDR CO line
intensities exceed those in PDRs in relatively weak radiation fields
($G_0=10^3$ or $F_X=1.6$~ergs~s$^{-1}$~cm$^{-2}$).  In XDRs with
stronger radiation fields, the lower $J$ CO lines (up to CO(4-3)) are
much weaker than in PDRs, but the higher rotational lines are
stronger.  All XDR CO ratios are (much) larger than even the high-CR
PDR CO ratios, and ratios at the intermediate and high $J$ levels
above CO(6-5) are diagnostically particularly meaningful.

{\it c. Density $n=10^5$~cm$^{-3}$} (Table \ref{high_dens}). There are
no longer significant differences between low-CR and high-CR PDRs.
However, the XDR CO lines are strong in the lower transitions, and
even more so at the intermediate and higher rotational transitions
above CO(4-3). Their ratios are always larger than the corresponding
ratios in any PDR model (see Meijerink et al. \citeyear{Meijerink2006}
for a more detailed discussion). Some high-$J$ CO intensities and
ratios are left blank, since no significant emission was found for
these lines due to the low fractional abundances of CO and the high
critical densities of the transitions.

\section{[CI]~609~$\mu$$\rm m$$/^{13}$CO(2-1) ratios}

XDRs, quite unlike PDRs, have significant neutral carbon abundances;
their [CI] intensities behave as volume tracers.  At the same time,
XDRs generally have weaker low J CO and $^{13}$CO lines than PDRs, for
$n\leq 10^4$~cm$^{-3}$ and $F_X>1$~ergs~s$^{-1}$~cm$^{-2}$. Thus, in
XDRs [CI]~609~$\mu$m/$^{13}$CO(2-1) line intensity ratios are much
larger than in corresponding PDRs.  Only at low densities, $n\leq
10^3$~cm$^{-3}$, do high-CR PDRs behave in a fashion intermediate
between XDRs and low-CR PDRs.  In such low-density, high-CR PDRs, CO
and $^{13}$CO dissociation causes simultaneous $^{13}$CO line
weakening and [CI] line strengthening, resulting in [CI]/$^{13}$CO
line intensity ratios four times higher than seen in low CR-PDRs, but
still (much) lower than those seen in XDRs.  Even at higher densities
($n\geq10^4$~cm$^{-3}$) and stronger radiation fields ($G_0=10^5$),
the high-CR PDRs fail to produce [CI]/$^{13}$CO ratios more than 1.5
times those of low-CR PDRs.  However, {\it in XDRs the [CI]/$^{13}$CO
ratios remain very much larger} and thus provide an excellent tool to
distinguish between PDRs and XDRs.

\section{HCN/CO and HCN/HCO$^+$ ratios}

\citet{Meijerink2006} found that HCN/CO and HCN/HCO$^+$ line intensity
ratios distinguish between `normal' (low-CR) PDRs and XDRs.  These
ratios are slightly different in high-CR PDRs.  At densities of
$n=10^4$~cm$^{-3}$, the HCN/CO ratios are lower in high-CR PDRs than
in low-CR PDRs, especially for the $J=4-3$ transition, but they become
more or less identical at densities of $n=10^5$~cm$^{-3}$.  In XDRs,
the HCN/CO ratios are almost invariably significantly lower than in either
PDR.  High-CR PDRs with densities of $n=10^4$~cm$^{-3}$ have
HCN/HCO$^+$ ratios {\it higher} than low-CR PDRs in the $J=1-0$
transition and {\it lower} than low-CR PDRs in the $J=4-3$
transition. This opposite behaviour reflects a cosmic ray induced
shift in the HCN and HCO$^+$ abundances to larger (cooler) columns.
However, at higher densities ($n=10^5$~cm$^{-3}$), high-CR PDRs have
HCN/HCO$^+$ ratios always lower than those in low-CR PDRs, mainly
because of a boost in the HCO$^+$ production. In all cases where XDRs
produce HCN and HCO$^{+}$ emission observable at all, the {\it
HCN/HCO$^+$ ratios are (much) lower in the XDR than in either PDR}.

\section{Conclusions}

\begin{enumerate}
\item CO line intensity ratios increase when cosmic ray ionization
rates are enhanced, but they remain smaller than those in XDRs. In
particular high-$J$ ($J>10$) CO lines (which will become observable
with HIFI in ESA's Herschel space observatory) allow to distinguish
between (high-CR) PDRs and XDRs. Using the HIFI time estimator and a
beam filling factor of 0.05, we find that a line intensity of
$9\times10^{-6}$~erg~$s^{-1}$~cm$^{-2}$, the largest CO(16-15)
intensity produced by our PDR models, will be detectable with HIFI in
about 4 hours, while the very bright lines received from highly
irradiated XDRs are detectable within minutes.

\item $[$CI$]$~609~$\mu$m/$^{13}$CO(2-1) ratios are much higher in
high-CR PDRs than in low-CR PDRs at modest densities of
$n=10^3$~cm$^{-3}$. At higher densities of $n\geq10^4$~cm$^{-3}$, this
difference vanishes. In XDRs, the ratios are always larger than in
PDRs at the same density, independent of incident radiation field.

\item HCN/CO and HCN/HCO$^+$ line ratios are good diagnostics to
distinguish between PDRs and XDRs.  As the ratios obtained for low-CR
and high-CR PDRs are different, combination with high-$J$ CO lines is
both crucial and profitable in the study of the active galaxy centers.

\item Although our model results broadly distinguish between low and
high CR PDRs and XDRs, there is some degeneracy when constraining the four
parameters (density, CR rate, $G_0$ and $F_X$) through individual ratios.
Therefore, a combination of various ratios should be considered.

\end{enumerate}

\begin{acknowledgements} 
We thank J.\ le Bourlot and M.\ Elitzur for
useful discussions on X-ray versus cosmic-ray
irradiation, M.\ Hogerheijde and P.\ van der Werf for discussions
on HIFI sensitivities, and an anonymous referee for his constructive
comments.
\end{acknowledgements}

\clearpage

\begin{table} 
\caption{PDR and XDR models}
\begin{center}
\begin{tabular}{cccc}
\tableline
\noalign{\smallskip}
Density  &  PDR$^1$ ($G_0$)           & XDR ($F_x$) & Size (pc)    \\ 
\noalign{\smallskip}
\hline
\noalign{\smallskip}
$10^3$     & $10^2$, $10^3$, $10^4$ & 0.16, 1.6, 16 & 10    \\
$10^4$     & $10^3$, $10^4$, $10^5$ & 1.6, 16, 160  & 1     \\
$10^5$     & $10^3$, $10^4$, $10^5$ & 1.6, 16, 160  & 1     \\
\noalign{\smallskip}
\tableline
\multicolumn{4}{l}{$^1$ Both low and high CR PDRs are calculated.}
\end{tabular}
\end{center}
\label{models}
\end{table}

\clearpage

\begin{table}
\caption{Line intensities (erg~s$^{-1}$~cm$^{-2}$~sr$^{-1}$) and ratios at density $n=10^3$~cm$^{-3}$}
\begin{flushleft}
{\scriptsize
\begin{tabular}{|c|ccc|ccc|ccc|}
\tableline
\noalign{\smallskip}
Model           &\multicolumn{3}{c}{Low-CR PDR}  &\multicolumn{3}{c}{High-CR PDR} & \multicolumn{3}{c}{XDR} \\             
Radiation field &$G_0=10^2$&$G_0=10^3$&$G_0=10^4$&$G_0=10^2$&$G_0=10^3$&$G_0=10^4$& $F_x=0.16$&$F_x=1.6$&$F_x=16$  \\
\noalign{\smallskip}
\tableline
\noalign{\smallskip}
CO(1-0)                       & 8.2(-8)  & 8.6(-8) & 1.0(-7) & 8.4(-8) & 8.3(-8) & 8.1(-8) & 1.2(-7) & 3.7(-10)& 3.0(-10) \\  
CO(2-1)                       & 4.8(-7)  & 5.1(-7) & 6.6(-7) & 6.3(-7) & 6.3(-7) & 6.4(-7) & 1.0(-6) & 9.3(-9) & 9.1(-9)  \\
CO(3-2)                       & 9.4(-7)  & 9.9(-7) & 1.5(-6) & 1.6(-6) & 1.6(-6) & 1.6(-6) & 2.8(-6) & 3.4(-8) & 5.2(-8)  \\
CO(4-3)                       & 1.4(-6)  & 1.6(-6) & 2.8(-6) & 2.2(-6) & 2.1(-6) & 2.1(-6) & 4.5(-6) & 5.3(-8) & 1.1(-7)  \\
CO(7-6)                       & 3.2(-9)  & 1.1(-7) & 7.8(-6) & 1.4(-7) & 1.2(-7) & 1.3(-7) & -       & 3.4(-8) & 1.1(-7)  \\
CO(10-9)                      & 8.4(-11) & 1.7(-11)& 2.3(-10)& 9.2(-9) & 4.7(-9) & 2.4(-9) & -       & -       & -        \\
CO(16-15)                     & -        & 1.0(-12)& 1.2(-10)& -       & 4.2(-9) & 7.1(-11)& -       & -       & -        \\
$^{13}$CO(2-1)                & 2.1(-7)  & 2.1(-7) & 2.8(-7) & 1.1(-7) & 1.0(-7) & 9.7(-8) & 8.1(-8) & 1.8(-10)& 1.6(-10) \\
$[$CI$]$ 609 $\mu$m           & 5.6(-6)  & 7.2(-6) & 8.9(-6) & 1.1(-5) & 1.2(-5) & 1.4(-5) & 3.6(-5) & 1.0(-4) & 1.2(-4)  \\
\noalign{\smallskip}
\tableline
\noalign{\smallskip}
CO(2-1)/CO(1-0)                    & 5.9      & 5.9     & 6.3     & 7.5     & 7.6     & 7.8     & 8.4     & 25.4    & 30.3   \\
CO(4-3)/CO(1-0)                    & 17.0     & 18.7    & 27.2    & 26.2    & 25.6    & 25.8    & 36.2    & 1.5(2)  & 3.6(2) \\ 
CO(7-6)/CO(3-2)                    & 3.4(-3)  & 1.1(-1) & 5.2     & 8.6(-2) & 7.9(-2) & 8.0(-2) & -       & 9.8(-1) & 2.2    \\
CO(10-9)/CO(7-6)                   & 2.5(-2)  & 1.5(-4) & 3.0(-5) & 6.7(-2) & 3.7(-2) & 1.9(-2) & -       & -       & -      \\
CO(16-15)/CO(1-0)                  & -        & 1.2(-5) & 1.1(-3) & -       & 5.0(-2) & 8.7(-4) & -       & -       & -      \\
CO(16-15)/CO(10-9)                 & -        & 6.2(-2) & 0.51    & -       & 9.0(-1) & 3.0(-2) & -       & -       & -      \\
$[$CI$]$ 609 $\mu$m/$^{13}$CO(2-1) & 26.7     & 34.1    & 31.5    & 97.1    & 1.2(2)  & 1.4(2)  & 4.4(2)  & 5.6(5)  & 7.4(5) \\
\noalign{\smallskip} 
\tableline
\noalign{\smallskip} 
\end{tabular}
}
\end{flushleft}
\label{low_dens}
\end{table}

\clearpage

\begin{table}
\caption{Line intensities (erg~s$^{-1}$~cm$^{-2}$~sr$^{-1}$) and ratios at density $n=10^4$~cm$^{-3}$}
\begin{flushleft}
{\scriptsize
\begin{tabular}{|c|ccc|ccc|ccc|}
\tableline
\noalign{\smallskip}
Model           &\multicolumn{3}{c}{Low-CR PDR} &\multicolumn{3}{c}{High-CR PDR}  & \multicolumn{3}{c}{XDR} \\             
Radiation field &$G_0=10^3$&$G_0=10^4$&$G_0=10^5$&$G_0=10^3$&$G_0=10^4$&$G_0=10^5$& $F_x=1.6$&$F_x=16$&$F_x=160$  \\
\noalign{\smallskip}
\tableline
\noalign{\smallskip}
CO(1-0)                       & 1.6(-7)  & 1.9(-7) & 2.5(-7) & 2.1(-7)  & 2.4(-7) & 2.7(-7) & 2.6(-7)  & 1.4(-10) & 3.4(-11) \\
CO(2-1)                       & 1.2(-6)  & 1.6(-6) & 2.0(-6) & 1.8(-6)  & 2.0(-6) & 2.4(-6) & 3.5(-6)  & 5.2(-9)  & 1.4(-9) \\
CO(3-2)                       & 3.4(-6)  & 4.6(-6) & 6.2(-6) & 5.3(-6)  & 6.2(-6) & 7.5(-6) & 1.3(-5)  & 4.0(-8)  & 1.0(-8) \\
CO(4-3)                       & 5.7(-6)  & 8.5(-6) & 1.2(-5) & 1.0(-5)  & 1.2(-5) & 1.5(-5) & 2.9(-5)  & 1.7(-7)  & 4.3(-8) \\
CO(7-6)                       & 2.9(-6)  & 1.3(-5) & 2.7(-5) & 1.6(-5)  & 2.3(-5) & 3.5(-5) & 9.7(-5)  & 1.8(-6)  & 6.4(-7) \\
CO(10-9)                      & 2.4(-9)  & 7.4(-8) & 1.8(-6) & 2.4(-7)  & 1.0(-6) & 4.9(-6) & 1.1(-4)  & 3.5(-6)  & 1.6(-6) \\
CO(16-15)                     & 1.1(-10) & 6.8(-9) & 2.3(-8) & 2.1(-10) & 3.1(-9) & 2.9(-8) & 1.1(-7)  & 2.5(-6)  & 1.7(-6) \\
HCN(1-0)                      & 5.2(-9)  & 6.8(-9) & 3.8(-9) & 3.9(-9)  & 3.8(-9) & 3.8(-9) & 4.9(-10) & 2.2(-13) & 3.1(-14)\\
HCN(4-3)                      & 3.3(-7)  & 4.0(-7) & 1.0(-8) & 9.5(-9)  & 1.0(-8) & 1.0(-8) & 1.3(-9)  & 4.0(-11) & 3.2(-11)\\
HCO$^+$(1-0)                  & 1.9(-8)  & 2.2(-8) & 2.6(-8) & 1.1(-8)  & 1.1(-8) & 1.1(-8) & 1.5(-8)  & 1.0(-14) & 1.4(-15)\\
HCO$^+$(4-3)                  & 4.3(-7)  & 6.6(-7) & 7.9(-7) & 6.2(-8)  & 7.0(-8) & 7.2(-8) & 2.0(-7)  & 1.0(-11) & 1.4(-12)\\
$^{13}$CO(2-1)                & 5.8(-7)  & 7.4(-7) & 9.3(-7) & 8.0(-7)  & 8.8(-7) & 9.6(-7) & 3.0(-7)  & 1.0(-10) & 2.3(-11)\\    
$[$CI$]$ 609 $\mu$m           & 8.4(-6)  & 9.9(-6) & 1.1(-5) & 1.2(-5)  & 1.4(-5) & 1.7(-5) & 5.2(-5)  & 1.2(-4)  & 1.2(-4) \\    
\noalign{\smallskip}
\tableline
\noalign{\smallskip}
CO(2-1)/CO(1-0)                    & 7.8      & 8.1     & 8.2     & 8.4      & 8.6     & 8.8     & 13.3     & 36.9     & 39.6     \\    
CO(4-3)/CO(1-0)                    & 35.8     & 44.2    & 50.4    & 47.5     & 51.8    & 56.9    & 1.1(2)   & 1.2(3)   & 1.3(3)   \\    
CO(7-6)/CO(3-2)                    & 8.4(-1)  & 2.8     & 4.3     & 3.1      & 3.7     & 4.7     & 7.5      & 46.1     & 61.9     \\    
CO(10-9)/CO(7-6)                   & 8.3(-4)  & 5.6(-3) & 6.5(-2) & 1.5(-2)  & 4.4(-2) & 1.4(-1) & 1.1      & 1.9      & 2.6      \\    
CO(16-15)/CO(1-0)                  & 7.0(-4)  & 3.5(-2) & 9.2(-2) & 1.0(-3)  & 1.3(-2) & 1.1(-1) & 4.1(-1)  & 1.8(4)   & 5.0(4)   \\    
CO(16-15)/CO(10-9)                 & 4.6(-2)  & 9.2(-2) & 1.2(-2) & 8.8(-4)  & 3.1(-3) & 6.0(-3) & 1.0(-3)  & 7.0(-1)  & 1.0      \\ 
HCN(1-0)/CO(1-0)                   & 3.2(-2)  & 3.5(-2) & 3.4(-2) & 1.9(-2)  & 1.6(-2) & 1.4(-2) & 1.9(-3)  & 1.6(-3)  & 9.1(-4)  \\     
HCN(4-3)/CO(4-3)                   & 5.7(-2)  & 4.7(-2) & 3.4(-2) & 9.5(-4)  & 8.4(-4) & 6.8(-4) & 4.4(-5)  & 2.4(-4)  & 7.3(-4)  \\    
HCN(1-0)/HCO$^+$(1-0)              & 2.8(-1)  & 3.1(-1) & 3.3(-1) & 3.6(-1)  & 3.4(-1) & 3.6(-1) & 3.2(-2)  & 21.7     & 22.6     \\    
HCN(4-3)/HCO$^+$(4-3)              & 7.6(-1)  & 6.1(-1) & 5.4(-1) & 1.5(-1)  & 1.5(-1) & 1.4(-1) & 6.3(-3)  & 4.0      & 22.6     \\
$[$CI$]$ 609 $\mu$m/$^{13}$CO(2-1) & 14.6     & 13.4    & 12.1    & 14.9     & 16.2    & 17.6    & 1.7(2)   & 1.2(6)   & 5.1(6)   \\  
\noalign{\smallskip} 
\tableline
\noalign{\smallskip} 
\end{tabular}
}
\end{flushleft}
\label{mid_dens}
\end{table}

\clearpage

\begin{table}
\caption{Line intensities (erg~s$^{-1}$~cm$^{-2}$~sr$^{-1}$) and ratios at density $n=10^5$~cm$^{-3}$}
\begin{flushleft}
{\scriptsize
\begin{tabular}{|c|ccc|ccc|ccc|}
\tableline
\noalign{\smallskip}
Model           &\multicolumn{3}{c}{Low-CR PDR}  &\multicolumn{3}{c}{High-CR PDR} & \multicolumn{3}{c}{XDR} \\             
Radiation field &$G_0=10^3$&$G_0=10^4$&$G_0=10^5$&$G_0=10^3$&$G_0=10^4$&$G_0=10^5$& $F_x=1.6$&$F_x=16$&$F_x=160$  \\
\noalign{\smallskip}
\tableline
\noalign{\smallskip}
CO(1-0)                       & 3.1(-7)  & 4.1(-7) & 5.5(-7) & 3.3(-7)  & 4.3(-7) & 5.6(-7) & 7.6(-7)  & 1.4(-6)  & 1.2(-6) \\
CO(2-1)                       & 2.7(-6)  & 3.5(-6) & 4.7(-6) & 2.8(-6)  & 3.7(-6) & 4.9(-6) & 7.1(-6)  & 1.4(-5)  & 1.5(-5) \\
CO(3-2)                       & 8.5(-6)  & 1.2(-5) & 1.6(-5) & 9.1(-6)  & 1.2(-5) & 1.7(-5) & 2.5(-5)  & 5.2(-5)  & 6.2(-5) \\
CO(4-3)                       & 1.8(-5)  & 2.5(-5) & 3.6(-5) & 1.9(-5)  & 2.7(-5) & 3.7(-5) & 5.6(-5)  & 1.2(-4)  & 1.7(-4) \\
CO(7-6)                       & 3.6(-5)  & 8.0(-5) & 1.4(-4) & 4.4(-5)  & 8.8(-5) & 1.4(-4) & 2.1(-4)  & 6.6(-4)  & 1.0(-3) \\
CO(10-9)                      & 4.3(-6)  & 7.7(-5) & 2.4(-4) & 1.2(-5)  & 9.0(-5) & 2.7(-4) & 3.7(-4)  & 1.6(-3)  & 2.9(-3) \\
CO(16-15)                     & 2.1(-8)  & 7.8(-7) & 5.1(-6) & 2.2(-8)  & 7.9(-7) & 8.8(-6) & 5.1(-6)  & 4.3(-3)  & 8.6(-3) \\
HCN(1-0)                      & 7.6(-8)  & 1.2(-7) & 1.5(-7) & 8.2(-8)  & 1.2(-7) & 1.5(-7) & 4.0(-8)  & 7.2(-8)  & 1.7(-8) \\
HCN(4-3)                      & 2.5(-6)  & 3.2(-6) & 3.7(-6) & 2.6(-6)  & 3.2(-6) & 3.7(-6) & 1.4(-6)  & 2.2(-6)  & 7.0(-7) \\
HCO$^+$(1-0)                  & 5.8(-8)  & 8.4(-8) & 1.1(-7) & 9.6(-8)  & 1.5(-7) & 2.3(-7) & 1.6(-7)  & 5.2(-7)  & 1.7(-7) \\
HCO$^+$(4-3)                  & 1.8(-6)  & 2.9(-6) & 4.0(-6) & 3.3(-6)  & 5.0(-6) & 6.8(-6) & 4.2(-6)  & 8.9(-6)  & 9.8(-6) \\
$^{13}$CO(2-1)                & 1.3(-6)  & 1.9(-6) & 4.7(-6) & 1.4(-6)  & 2.0(-6) & 2.6(-6) & 2.8(-6)  & 4.9(-6)  & 2.3(-6) \\
$[$CI$]$ 609 $\mu$m           & 7.4(-6)  & 8.7(-6) & 9.8(-6) & 9.1(-6)  & 1.1(-5) & 1.3(-5) & 4.1(-5)  & 1.1(-4)  & 4.4(-4) \\
\noalign{\smallskip}
\tableline
\noalign{\smallskip}
CO(2-1)/CO(1-0)                    & 8.6      & 8.5     & 8.5     & 8.6      & 8.6     & 8.7     & 9.3      & 10.0     & 12.2    \\
CO(4-3)/CO(1-0)                    & 56.6     & 61.1    & 64.6    & 57.8     & 62.3    & 66.2    & 73.1     & 91.1     & 1.3(2)  \\
CO(7-6)/CO(3-2)                    & 4.2      & 6.9     & 9.1     & 4.8      & 7.14    & 8.8     & 8.9      & 12.9     & 16.5    \\
CO(10-9)/CO(7-6)                   & 1.2(-1)  & 1.0     & 1.7     & 2.6(-1)  & 1.0     & 1.8     & 1.7      & 2.4      & 2.8     \\
CO(16-15)/CO(1-0)                  & 6.6(-2)  & 1.9     & 9.2     & 6.7(-2)  & 1.8     & 15.6    & 6.7      & 3.1(3)   & 6.9(3)  \\
CO(16-15)/CO(10-9)                 & 4.8(-3)  & 1.0(-2) & 2.1(-2) & 1.9(-3)  & 8.7(-3) & 3.3(-2) & 1.4(-2)  & 2.7      & 3.0     \\
HCN(1-0)/CO(1-0)                   & 2.4(-1)  & 2.8(-1) & 2.8(-1) & 2.5(-1)  & 2.7(-1) & 2.7(-1) & 5.2(-2)  & 5.1(-2)  & 1.3(-2) \\
HCN(4-3)/CO(4-3)                   & 1.4(-1)  & 1.2(-1) & 1.0(-1) & 1.3(-1)  & 1.1(-1) & 1.0(-1) & 2.5(-2)  & 1.7(-2)  & 4.2(-3) \\
HCN(1-0)/HCO$^+$(1-0)              & 1.3      & 1.4     & 1.4     & 8.5(-1)  & 8.0(-1) & 6.7(-1) & 2.5(-1)  & 1.4(-1)  & 9.9(-2) \\
HCN(4-3)/HCO$^+$(4-3)              & 1.4      & 1.1     & 9.3(-1) & 7.8(-1)  & 6.3(-1) & 5.4(-1) & 3.2(-1)  & 2.5(-1)  & 7.1(-2) \\
$[$CI$]$ 609 $\mu$m/$^{13}$CO(2-1) & 5.7      & 4.5     & 3.6     & 6.3      & 5.5     & 4.8     & 14.7     & 24.0     & 1.9(2)  \\
\noalign{\smallskip} 
\tableline
\noalign{\smallskip} 
\end{tabular}
}
\end{flushleft}
\label{high_dens}
\end{table}

\clearpage

\end{document}